\documentclass[reprint,nofootinbib,amsmath,amssymb,aps,prd,floatfix,]{revtex4-2}

\usepackage{graphicx}
\usepackage{dcolumn}
\usepackage{bm}
\usepackage[caption=false]{subfig}
\usepackage{tabularx}
\usepackage{float}
\usepackage[separate-uncertainty=true,multi-part-units=single]{siunitx}
\usepackage{listings}
\usepackage{hyperref}
\usepackage{cleveref}
\usepackage{amsmath}
\usepackage{caption}
\usepackage{subcaption}
\usepackage{braket}
\usepackage{tensor}
\usepackage{tabularx}
\usepackage{booktabs}
\usepackage[scr=boondox]{mathalpha}
\usepackage{amssymb}
\usepackage{pifont}
\usepackage[nolist,nohyperlinks]{acronym}
\usepackage{enumerate}
\usepackage{orcidlink}
\DeclareSIUnit \parsec {pc}
\DeclareSIUnit \year{yr}

\newcommand{\crefrangeconjunction}{--}
\newcommand*\Eval[3]{\left.#1\right\rvert_{#2}^{#3}}

\crefrangelabelformat{equation}{(#3#1#4 \crefrangeconjunction ~#5#2#6)}
\DeclareSIUnit\pc{pc}

\begin{document}

\crefname{section}{Sec.}{Sec.}
\crefname{equation}{Eq.}{Eqs.}
\crefname{figure}{Fig.}{Fig.}
\crefname{table}{Table}{Table}
\crefname{appendix}{Appx.}{Appx.}

\title{Constraining the Locally Rotationally Symmetric Bianchi Type I Model with Self-Consistent Recombination History and Observables}

\author{Boris Hoi-Lun \surname{Ng}~\orcidlink{0009-0000-4811-5294}}
\email{boris.hl.ng@link.cuhk.edu.hk}
\affiliation{%
Department of Physics, The Chinese University of Hong Kong, Shatin, Hong Kong
}%
\author{Ming-Chung \surname{Chu}~\orcidlink{0000-0002-1971-0403}}
\email{mcchu@phy.cuhk.edu.hk}
\affiliation{%
Department of Physics, The Chinese University of Hong Kong, Shatin, Hong Kong
}%
\date{\today}
\begin{abstract}
Recent cosmological measurements suggest the possibility of an anisotropic universe. As a result, the Bianchi Type I metric, being the simplest anisotropic extension to the standard Friedmann–Lemaître–Robertson–Walker metric has been extensively studied. In this work, we show how the recombination history should be modified in an anisotropic universe and derive observables by considering the null geodesic. We then constrain the Locally Rotationally Symmetric Bianchi Type I model by performing Markov Chain Monte Carlo with the acoustic scales in Cosmic Microwave Background (CMB) and Baryon Acoustic Oscillation data, together with local measurements of $H(z)$ and Pantheon Supernova data. Our results reveal that the anisotropic model is not statistically preferred over the $\Lambda$CDM model, and we obtain a tight constraint on the anisotropy that generally agrees with previous studies under a maximum temperature anisotropy fraction of \num{2e-5}. Besides, we also present constraints based on a relaxed maximum temperature anisotropy. We stress that there is a significant difference between the geodesic-based observables and the naive isotropic analogies when there is a noticeable anisotropy. However, the changes in recombination history are insignificant even under the relaxed anisotropy limit.
\end{abstract}
\maketitle

\begin{acronym}
    \acro{CP}{Cosmological Principle}
    \acro{FLRW}{Friedmann–Lemaître–Robertson–Walker}
    \acro{CMB}{Cosmic Microwave Background}
    \acro{WMAP}{Wilkinson Microwave Anisotropy Probe}
    \acro{BAO}{Baryon Acoustic Oscillation}
    \acro{MCMC}{Markov Chain Monte Carlo}
    \acro{SNIa}{Type Ia Supernova}
    \acroplural{SNIa}[SNeIa]{Type Ia Supernovae}
    \acro{BBN}{Big Bang nucleosynthesis}
    \acro{AIC}{Akaike information criterion}
    \acro{CIB}{Cosmic Infrared Background}
    \acro{EOS}{equation of state}
    \acro{LRS}{Locally Rotationally Symmetric}
\end{acronym}
\section{\label{Intro}Introduction}
The \ac{CP} of homogeneity and isotropy on large scales is a foundational assumption in the standard $\Lambda$CDM cosmological model with the flat \ac{FLRW} metric
\begin{equation}
    ds^2=-dt^2 + \mathfrak{a}^2(t) \left( dx^2 + dy^2 +dz^2\right),
\end{equation}
where $\mathfrak{a}$ is the usual scale factor. However, recent analyses of cosmological measurements at both large and local scales challenge this assumption.

Global measurements, such as those of the \ac{CMB}, have long been suspected to exhibit large-scale anisotropies. Even back in the \textit{\ac{WMAP}} data, anomalies that hint at a large-scale spatial anisotropy emerge, such as the alignment of the quadrupole and octopole axes \cite{Bielewicz_2005, deOliveira-Costa_2004, Tegmark_2003, Wiaux_2006}, the North-South asymmetry \cite{Hansen_2004, Eriksen_2004, Bernui_2006, Bernui_2008}, and hints of some preferred axes \cite{Bunn_2000, Hansen_2004, Eriksen_2004, Bernui_2007, Hansen_2009}. Furthermore, several cosmological parameters ($n_s$, $A_s$, and $\Omega_{b0}$) exhibit anisotropic distributions across the full sky \cite{Axelsson_2013}.

Such anomalies did not disappear with the more precise data from \textit{Planck}. Instead, more and more evidence suggests the existence of large-scale anisotropy. An example is the \ac{CMB} parity asymmetry which has a preferred axis roughly aligning the \ac{CMB} dipole, quadrupole, and octopole \cite{Naselsky_2012, Zhao_2014, Cheng_2016, Aluri_2017}. In addition, all six cosmological parameters $\set{\Omega_{b} h^2, \Omega_{c} h^2, n_s, 100\theta_\mathrm{MC}, \tau, \ln{(10^{10}A_s)}}$, in addition to the Hubble constant $H_0$, are found to exhibit a dipole distribution at 2-3$\sigma$ level, with generally aligned dipole directions \cite{Yeung_2022}. Hence, both the \acfi{WMAP} and \textit{Planck} data indicate an anisotropic universe, contradicting the assumed \ac{CP}.

Interestingly, several local measurements also show anisotropic features. In particular, the bulk flow of galaxies \cite{Kashlinsky_2008, Watkins_2009} and the quasar polarization \cite{Hutsemekers_2005} have been observed to align with the \ac{CMB} dipole. Various \ac{SNIa} measurements \cite{Antoniou_2010, Javanmardi_2015, Krishnan_2022, Luongo_2022, Hu_2023} also suggest directional dependency in cosmic acceleration and $H_0$, with a generally larger value near the \ac{CMB} dipole direction \cite{Javanmardi_2015, Krishnan_2022, Luongo_2022, Hu_2023}. However, the directions of the preferred axes vary in different studies. For a detailed review, see Ref. \cite{Antoniou_2010, Zhao_2016, Perivolaropoulos_2022}.

These observations suggest that either significant unknown systematics pollute both the \ac{CMB} and local measurements or \ac{CP} is indeed inexact. We are thus motivated to consider a Bianchi Type I metric 
\begin{gather}
    ds^2 = -dt^2 + a^2(t)dx^2 + b^2(t)dy^2 + c^2(t)dz^2, \label{Bianchi metric}
\end{gather}
which is anisotropic but homogeneous and is a natural anisotropic extension of the \ac{FLRW} metric \cite{Akarsu_2019,Shekh_2020,Sarmah_2022,Koussour_2023, Akarsu_2023, Hertzberg_Loeb_2024}. Here, $a(t)$, $b(t)$, and $c(t)$ are the scale factors in the $x$, $y$ and $z$ directions, respectively, which may be distinct from the \ac{FLRW} scale factor $\mathfrak{a}$. To reduce the number of parameters, we consider a subset of the Bianchi Type I spacetime extension of the $\Lambda$CDM model (Bianchi Type I model), the \ac{LRS} model, where $a(t)=b(t)$, though our method can be extended to the general Bianchi Type I model.

There are works that constrain the degree of anisotropy of a Bianchi Type I extended universe using global and local cosmological data \cite{Akarsu_2019, Shekh_2020, Sarmah_2022, Yadav_2023, Koussour_2023, Akarsu_2023, Koivisto_2008}. The authors of Ref. \cite{Akarsu_2019,Sarmah_2022, Yadav_2023, Koussour_2023, Akarsu_2023} took a phenomenological approach, treating the anisotropy as a stiff fluid while considering the evolution of the geometric mean of the scale factors $a,\ b$ and $c$ by a modified Friedmann equation and observables constructed from the mean scale factors. Besides, some authors would fix the physical baryon density to the $\Lambda$CDM value \cite{Akarsu_2019,Yadav_2023}. Meanwhile, although Ref. \cite{Hertzberg_Loeb_2024} did not adopt such definitions of observables, only order-of-magnitude constraints were derived at the fixed decoupling time $t_\mathrm{CMB}\sim\SI{0.38}{\mega\year}$, based on the \ac{CMB} temperature anisotropy or at the fixed time of \ac{BBN}. These prompt us to re-constrain an \ac{LRS} Bianchi Type I model with self-consistently defined recombination history, distances, and hence observables.

In \cref{Model}, we first introduce the anisotropic universe model considered. In particular, the expansion equations and corresponding thermodynamics history are discussed. Subsequently, we review the definitions of observables adopted by previous works in \cref{Distance_Observables} and present how comoving distances, observables and their full-sky averages should be defined.  Next, we discuss how we constrain the anisotropic cosmological parameters (\cref{Method_data}). The results are presented and discussed in \cref{Results}. Finally, in \cref{Conclusion}, we conclude.

\section{The Anisotropic universe \label{Model}}
\subsection{LRS Bianchi Type I Model \label{GR}}
In the Bianchi Type I model, the line element in the matter-comoving coordinate (4-velocity $u^\mu=\delta^\mu_0$) is given by \cref{Bianchi metric} \footnote{Throughout this paper, we adopt the metric signature $(-,+,+,+)$ and natural units where speed of light and gravitational constant are unity ($c=G=1$).} \cite{Akarsu_2019,Yadav_2023,Hertzberg_Loeb_2024,Sarmah_2022,Akarsu_2023,Koivisto_2008}.
Note that the corresponding scale factors for the 3 directions $\set{a(t), b(t), c(t)}$ are generally different, implying an anisotropic universe. Subsequently, the Hubble parameters in the 3 directions are defined as
\begin{align}
    H_x&=\frac{\dot{a}}{a}, &
    H_y&=\frac{\dot{b}}{b}, &
    H_z&=\frac{\dot{c}}{c}.
\end{align}
The dot here denotes a derivative with respect to cosmic time $t$.

On the other hand, the energy-momentum tensor $T\indices{^\mu_{\nu}}$, including radiation ($r$), matter ($m$), and dark energy (in the form of the cosmological constant $\Lambda$) takes the form
\begin{equation}
\begin{aligned}
    T\indices{^\mu_{\nu}} &= \text{diag}(-\rho, p_x, p_y, p_z),\\
    &=\text{diag}(-\rho, p, p, p)
\end{aligned}
\end{equation}
where $\rho=\sum \rho_X = \rho_r+\rho_m+\rho_\Lambda$ and $p=\sum p_X =p_r+p_m+p_\Lambda$ are, respectively, the total energy density and the total pressure. Note that the assumption of isotropic pressure is applied in the second line \cite{Akarsu_2019,Sarmah_2022,Yadav_2023,Shekh_2020,Akarsu_2023,Hertzberg_Loeb_2024} and radiation includes contribution from massless neutrinos. Applying the local conservation law
\begin{align}
    \nabla_\mu T\indices{^{\mu\nu}}=0
\end{align}
to the energy-momentum tensor as well as assuming that the species are non-interacting and minimally coupled gives the time evolution equation
\begin{align}
    \dot{\rho}_X+(1+w_X)\rho_X(H_x+H_y+H_z)&=0 \label{d_rho},
\end{align}
with $p_X=w_X\rho_X$. Here, the \ac{EOS} parameters $(w_r, w_m, w_\Lambda)$ equal to $(1/3, 0, -1)$. To study how energy densities vary with the scale factors, \cref{d_rho} can be solved analytically
\begin{gather}
    \rho_X = \rho_{X0}\left(\frac{abc}{a_0 b_0 c_0}\right)^{-(1+w_X)}.\label{rho_scale}
\end{gather}
The subscript `0' denotes the current-day value throughout this work. The energy densities scale just like under the \ac{FLRW} metric but with $\mathfrak{a}^3$ replaced by $(abc)$.

The expansion history is given by solving the Einstein Field Equation
\begin{align}
    G\indices{_{\mu\nu}}=8\pi T\indices{_{\mu\nu}},
\end{align}
which contains 4 non-trivial equations
\begin{gather}
    H_x H_y + H_y H_z + H_z H_x = 8\pi\rho \label{EFE1},\\
    \frac{\ddot{b}}{b}+ H_y H_z + \frac{\ddot{c}}{c} = -8\pi p \label{EFE2},\\
    \frac{\ddot{c}}{c}+ H_z H_x + \frac{\ddot{a}}{a} = -8\pi p \label{EFE3},\\
    \frac{\ddot{a}}{a}+ H_x H_y + \frac{\ddot{b}}{b} = -8\pi p \label{EFE4}.
\end{gather}
Note that \cref{EFE1} acts like a Friedmann equation.

Now, we restrict ourselves to an \ac{LRS} model $(a=b)$ \cite{Hertzberg_Loeb_2024,Koussour_2023} and examine \cref{EFE2,EFE3,EFE4}, which reduce to
\begin{gather}
    \ddot{a} = -a\left(4\pi p +\frac{1}{2} H_x^2\right)  \label{dd_a},\\
    \ddot{c} = - c \left(8\pi p + H_xH_z + \frac{\ddot{a}}{a} \right) \label{dd_c}.
\end{gather}
Combining \cref{dd_a,dd_c} with the \ac{LRS} version of \cref{d_rho}
\begin{gather}
    \dot{\rho}_X+(1+w_X)\rho_X(2H_x+H_z)=0, \label{d_rho_axial}
\end{gather}
one can solve the time evolution of the universe by providing the initial conditions $(H_{x0}, H_{z0},\rho_{m0}, \rho_{r0}, \rho_{\Lambda0})$. In particular, the current values of the scale factors are set to one, i.e., $a_0=c_0=1$ since the line element in \cref{Bianchi metric} is invariant if $a\rightarrow \lambda a,\ dx\rightarrow dx/\lambda$ and similarly for $c$. Lastly, it is important to emphasize that we are considering a flat universe, $ \rho_{r0} + \rho_{m0} + \rho_{\Lambda0} = \rho_\text{crit0}$, the critical density of the universe. Hence, we only have 2 energy densities as model parameters. From \cref{EFE1}, the flatness criterion implies \cite{Russell_2014} 
\begin{gather}
    \rho_{r0} + \rho_{m0} + \rho_{\Lambda0} = \rho_{\text{crit}0} = \frac{H_{x0}^2 + 2 H_{x0} H_{z0}}{8\pi}.
\end{gather}

\subsection{Thermodynamics of the Anisotropic Universe \label{Thermo}}
One needs to consider the thermodynamics of the universe to track the recombination history. This work assumes thermal equilibrium from the beginning of time till the last scattering of \ac{CMB} photons, justified by the rapid interactions. After the last scattering, photons propagate freely while maintaining the shape of a blackbody spectrum. Under the Bianchi Type I model, photons traveling in different directions experience different cosmological redshifts due to the directional-dependent scale factor. Hence, the temperature and the blackbody spectrum would have a directional dependency. However, by enforcing that the \ac{CMB} temperature anisotropies at present-day are of the order of \num{e-5} \cite{Planck_2018}, the directional dependency of temperature is negligible, and an isotropic temperature is a good approximation. Furthermore, we stress that the anisotropy exists only at the metric level, manifesting through the scale factors. Hence, the usual thermodynamics equations are valid with suitable modifications related to the scale factors. For instance, by considering radiation with $\rho_r\propto T^4$ (from statistical mechanics) and $\rho_r\propto (abc)^{-4/3}$ [from \cref{rho_scale}], we obtain
\begin{align}
    T\propto (abc)^{-1/3},\label{Tempearture}
\end{align}
where $T$ is the \ac{CMB} temperature, analogous to that under the \ac{FLRW} metric.

The thermodynamics are crucial in finding the moment of the last scattering. There are well-developed thermodynamics codes such as \texttt{HyRec} \cite{HyRec-2, HyRec} that calculate the recombination process by keeping track of complicated atomic transitions. However, they use the \ac{FLRW} scale factor $\mathfrak{a}$ as the temporal coordinate and thus cannot be easily modified for our model. Instead, we consider the contribution from Hydrogen only and follow Peebles' formalism \cite{Peebles_1968,Baumann_2022,Weinberg_2008} to approximate the electron fraction $X_e$ evolution and the recombination history. This approach also offers the advantage of providing a more analytical understanding of how the recombination history is influenced by anisotropy. Most factors in the Peebles' formalism are independent of the expansion of the universe, so we need not change anything except for $\Lambda_\alpha$, which is the rate of recombination via resonance escape \cite{Baumann_2022}. Under the usual \ac{FLRW} metric,
\begin{gather}
    \Lambda_\alpha=\frac{27}{128\zeta(3)}\frac{\mathfrak{H}(t)}{[1-X_e(t)]\eta[k_B T(t)/ E_I]^3} ,\label{Lambda_alpha}
\end{gather}
where $\eta$ is the baryon-to-photon ratio, $k_B$ is the Boltzmann constant, and $E_I$ is the ionization energy of Hydrogen. Here, $\mathfrak{H}$ is the Hubble parameter under the FLRW metric that describes the cosmological redshift of photons and thus needs to be modified \cite{Baumann_2022,Weinberg_2008}. Suppose a photon is emitted in a \ac{FLRW} universe at $t=t_e$ and is measured at $t=t'$, the photon frequency is redshifted by 
\begin{align}
    \frac{\Eval{\mathfrak{a}}{t_e}{}}{\Eval{\mathfrak{a}}{t'}{}}.
\end{align}
By Taylor expanding near $t'=t_e$, we have
\begin{gather}
    \frac{\Eval{\mathfrak{a}}{t_e}{}}{\Eval{\mathfrak{a}}{t'}{}} = 1- \Eval{\frac{\dot{\mathfrak{a}}}{\mathfrak{a}}}{t_e}{}(t'-t_e)= 1- \mathfrak{H}(t_e)(t'-t_e).
\end{gather}
For an \ac{LRS} Bianchi Type I metric, the photon frequency is redshifted by
\begin{align}
    \frac{\Eval{(a^2 c)^{1/3}}{t_e}{}}{\Eval{(a^2 c)^{1/3}}{t'}{}}.
\end{align}
Expanding around $t'=t_e$, we obtain
\begin{gather}
     \frac{\Eval{(a^2 c)^{1/3}}{t_e}{}}{\Eval{(a^2 c)^{1/3}}{t'}{}} = 1- \frac{1}{3}\left[2\frac{\dot{a}}{a} + \frac{\dot{c}}{c}\right]_{t_e}(t'-t_e).
\end{gather}
Therefore, by replacing 
\begin{gather}
    \mathfrak{H} \to \frac{1}{3}\left(2\frac{\dot{a}}{a} + \frac{\dot{c}}{c}\right),
\end{gather}
one obtains the resonance escape rate $\Lambda_\alpha$ under the \ac{LRS} universe and by extension, the $X_e$ evolution.

Consequently, we are equipped to find the optical depths by ignoring the contribution from reionization. Following the definition of \textit{Planck} \cite{Planck_2013,Planck_2015,Planck_2018}, we denote the time when the photon optical depth
\begin{gather}
    \tau=\int_{t_*}^{t_0} \sigma_T n_b X_e dt \label{tau}
\end{gather}
and the time when the baryon-drag optical depth
\begin{gather}
    \tau_\mathrm{drag}=\int_{t_\mathrm{drag}}^{t_0} R\sigma_T n_b X_e dt \label{tau_drag}
\end{gather}
equal to unity by $t_*$ and $t_\mathrm{drag}$, respectively. Note that $\sigma_T$ is the Thomson cross-section, $n_b$ is the baryon number density and $R= [3\rho_b(t)]/[4\rho_r(t)]$.

\section{Distance and Observables \label{Distance_Observables}}
\subsection{Comments on Previous Definitions}
Ref. \cite{Akarsu_2019,Sarmah_2022, Yadav_2023} solved the expansion history of a Bianchi Type I spacetime extension of the $\Lambda$CDM model phenomenologically by treating the anisotropy as an ideal stiff fluid and considering the evolution of the geometric mean of the directional scale factors
\begin{gather}
    \tilde{a} = \sqrt[3]{abc},\\
    1 + \tilde{z} = \frac{1}{\tilde{a}}, \label{Akarsu_redshift}\\
    \tilde{H} = \frac{\dot{\tilde{a}}}{\tilde{a}}=\frac{1}{3}(H_x + H_y + H_z).
\end{gather}
From this, they derived the corresponding first Friedmann equation \cite{Akarsu_2019,Sarmah_2022, Yadav_2023}
\begin{gather}
    \tilde{H}(\tilde{a})^2 = \tilde{H}_0^2 \left( \tilde{\Omega}_{r0}\tilde{a}^{-4} + \tilde{\Omega}_{m0}\tilde{a}^{-3} + \tilde{\Omega}_{\Lambda0} + \tilde{\Omega}_{\sigma0}\tilde{a}^{-6} \right),
\end{gather}
where $\tilde{\Omega}_{X0}$ here is the dimensionless energy density defined using $\tilde{H}$ instead of the $\Lambda$CDM $\mathfrak{H}$. Moreover, there is an additional energy density $\tilde{\Omega}_{\sigma0}=\sigma^2/(3\tilde{H}^2)$ generated by the shear scalar $\sigma^2$ \cite{Akarsu_2019,Sarmah_2022, Yadav_2023}, which captures the anisotropic effects. The above is equivalent to the previous treatment in \cref{GR} and can even be applied in \cref{Thermo}. The authors obtained a neat modified Friedmann equation in exchange for information on individual scale factors $a$, $b$, and $c$.

However, the phenomenological approach has limitations. Because there is only a single scale factor $\tilde{a}$ and anisotropy is encoded in the stiff fluid energy density $\tilde{\Omega}_\sigma$, all physical quantities remain isotropic. For example, the authors \cite{Akarsu_2019,Yadav_2023, Akarsu_2023} define the comoving sound horizon at drag redshift
\begin{gather}
    \tilde{r}_s = \int_{\tilde{z}_d}^\infty \frac{c_s d\tilde{z}}{\tilde{H}(\tilde{z})},
\end{gather}
in the same way as in the \ac{FLRW} metric. Clearly, the sound horizon $\tilde{r}_s$ lacks directional dependence, a key feature of anisotropic models. To move beyond isotropic observables, one must therefore go beyond the phenomenological approach and track individual scale factors.

\subsection{Comoving Distance \label{Distance}}
The comoving distance is defined as the distance between two points in comoving coordinates \cite{Delliou_2020}
\begin{gather}
    d D_C^2 = dx^2 + dy^2 + dz^2.
\end{gather}
Now, consider a null geodesic \cite{Delliou_2020}
\begin{gather}
    1 = a^2 \left(\frac{dx}{dt}\right)^2 + b^2 \left(\frac{dy}{dt}\right)^2 + c^2 \left(\frac{dz}{dt}\right)^2,
\end{gather}
connecting the observer who is, without loss of generality, positioned at the origin, and the source which is positioned at a polar angle $\Theta$ and an azimuthal angle $\phi$  such that
\begin{gather}
    \frac{dy}{dx} = \tan\phi,\\
    \frac{d\ell}{dz} = \tan\Theta,\\
    d\ell^2 = dx^2 + dy^2.
\end{gather}
Thus, in agreement with Ref. \cite{Koivisto_2008},
\begin{gather}
    D_C(\Theta, \phi, t_1, t_2) = \int_{t_1}^{t_2} \frac{dt}{\sqrt{\xi^2\sin^2\Theta +c^2 \cos^2\Theta}},\label{D_c_general}\\
    \xi^2 = a^2\cos^2\phi + b^2\sin^2\phi.
\end{gather}

For an \ac{LRS} universe ($a=b$), \cref{D_c_general} is simplified to
\begin{gather}
    D_C(\Theta, t_1, t_2) = \int_{t_1}^{t_2} \frac{dt}{\sqrt{a^2 \sin^2\Theta +c^2 \cos^2\Theta}} \label{comoving_distance}.
\end{gather}
An effective scale factor can thus be defined
\begin{gather}
    a_\mathrm{eff}(\Theta, t) = \sqrt{a(t)^2 \sin^2\Theta +c(t)^2 \cos^2\Theta},\label{effective_a}\\
\intertext{such that}
    D_C(\Theta, t_1, t_2) = \int_{t_1}^{t_2} \frac{dt}{a_\mathrm{eff}},
\end{gather}
resembling the comoving distance in the \ac{FLRW} metric, but now with explicit directional dependency.

\subsection{Observables \label{Observables}}
Observables like angular acoustic scales in \ac{CMB} and \ac{BAO} as well as Supernova distance moduli are derived from more fundamental quantities such as the transverse comoving diameter distance. In the following, we discuss how the observables are defined under the \ac{LRS} model.

\subsubsection{Directional Hubble Parameter}
The effective scale factor at the polar angle $\Theta$ and time $t_\mathrm{obs}$ is defined in \cref{effective_a}. Consequently, one can construct the corresponding Hubble parameter
\begin{equation}
\begin{aligned}
    H(\Theta, t_\mathrm{obs}) &= \frac{1}{a_\mathrm{eff}(\Theta, t_\mathrm{obs})}\frac{da_\mathrm{eff}}{dt}\\
                &= \frac{a \dot{a} \sin(\Theta)^2 + c \dot{c} \cos(\Theta)^2}{a^2 \sin(\Theta)^2 + c^2 \cos(\Theta)^2}.
\end{aligned}
\end{equation}

\subsubsection{Comoving Sound Horizon}
Considering a general last scattering time $t_\mathrm{ls}$, we derive the equation for the comoving sound horizon $r_\mathrm{ls}$. At $t_\mathrm{ls}$, the sound waves would have propagated out to an ellipsoidal surface. According to \cref{comoving_distance}, its semi-axes along the principal directions are
\begin{gather}
\begin{align}
        L_{x,y} &= \int_0^{t_\mathrm{ls}} \frac{c_s(t)dt}{a(t)}, && \text{for $x$ and $y$-direction}, \\
        L_{z} &= \int_0^{t_\mathrm{ls}} \frac{c_s(t)dt}{c(t)}, && \text{for $z$ direction},
\end{align}\\
        c_s(t) = \frac{1}{\sqrt{3(1+R(t))}}.
\end{gather}

\begin{figure}
    \centering
    \includegraphics[width=\columnwidth]{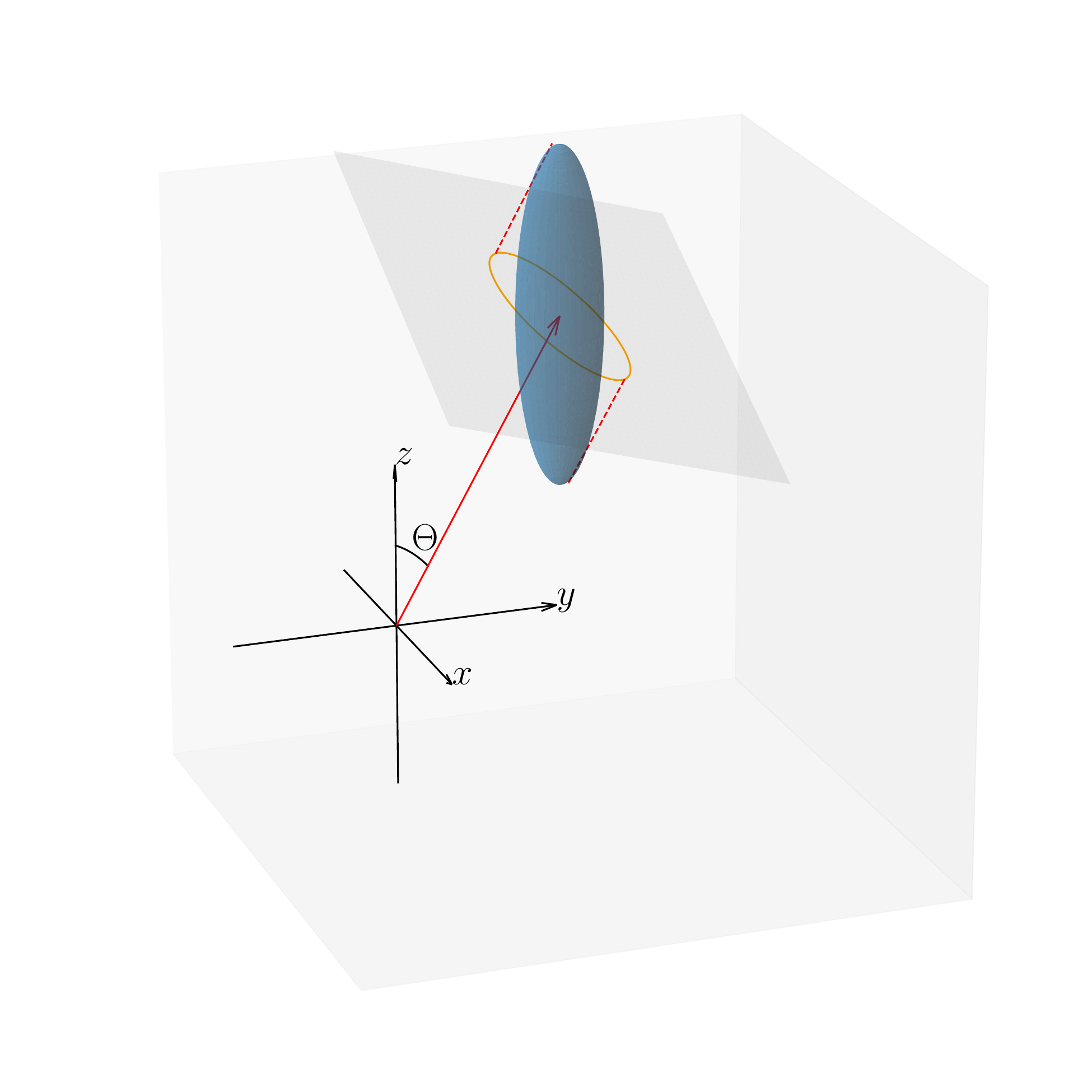}
    \caption{Schematic of the sound horizon at a polar angle $\Theta$. The sound waves form an ellipsoidal surface (the blue surface) and the observer at the origin would observe an ellipse (the orange line) as the projection along the radial direction.\label{ellipsoid_schematic}}
\end{figure}
The maximum region of correlation observed at a polar angle $\Theta$ is defined by the projection of the ellipsoid along the radial direction, as indicated by the orange ellipse in \cref{ellipsoid_schematic}. By geometry, an observer at the origin observes an ellipse with semi-axes
\begin{gather}
    \tilde{L}_1 = L_{x,y},\\
    \tilde{L}_2 = \sqrt{L_{x,y}^2 \cos^2\Theta+L_{z}^2 \sin^2\Theta}.
\end{gather}
Note that due to the symmetry in the $x$ and $y$-axes, $\tilde{L}_1=L_{x,y}$. We consider the area of the ellipse and define an effective comoving sound horizon by
\begin{gather}
    r_\mathrm{ls}(\Theta, t_{\rm ls}) = \sqrt{\frac{\textrm{Area}(\Theta, t_{\rm ls})}{\pi}},\\
    \mathrm{Area} = \pi \tilde{L}_1 \tilde{L}_2 = \pi L_{x,y} \sqrt{L_{x,y}^2 \cos^2\Theta+L_{z}^2 \sin^2\Theta}.
\end{gather}

\subsubsection{Angular Scales and Inverse Angular Scales \label{Angular_Scales}}
The \ac{CMB} angular acoustic scale at angle $\Theta$ is defined as the ratio of the sound horizon $r_*$ and the transverse comoving diameter distance $D_{M,*}$:
\begin{gather}
    \theta_*(\Theta) = \frac{r_*}{D_{M,*}(\Theta)},\\
    r_*(\Theta) = r_\mathrm{ls}(\Theta, t_*),\\
    D_{M,*}(\Theta) = D_C(\Theta, t_*, t_0).
\end{gather}

The \ac{BAO} observables are defined similarly. For objects at time $t_\mathrm{obs}$, we follow the definition of \textit{DESI} \cite{DESI_2024mwx} and define the \ac{BAO}-related distances as 
\begin{gather}
    r_d(\Theta) = r_\mathrm{ls}(\Theta, t_d), \label{BAO 1}\\
    D_{M,d}(\Theta, t_\mathrm{obs}) = D_C(\Theta,  t_\mathrm{obs}, t_0),\label{BAO 2}\\
    D_{A,d}(\Theta, t_\mathrm{obs}) = a_\mathrm{eff}(\Theta, t_\mathrm{obs}) D_{M,d}(\Theta, t_\mathrm{obs}),\label{BAO 3}\\
    D_{H,d}(\Theta, t_\mathrm{obs}) = \frac{1}{H(\Theta, t_\mathrm{obs})},\label{BAO 4}\\
    D_{V,d}(\Theta, t_\mathrm{obs}) = \left(D_{M,d}^2 \frac{1}{a_\mathrm{eff}+1}D_{H,d}\right)^{1/3}.\label{BAO 5}
\end{gather}
From \cref{BAO 1,BAO 2,BAO 3,BAO 4,BAO 5}, we can construct the (inverse) angular acoustic scales $D_{M,d}/r_d$, $D_{A,d}/r_d$, $D_{H,d}/r_d$, $D_{V,d}/r_d$, $H r_d$ and $r_d/D_{V,d}$.

\subsubsection{Type Ia Supernova Luminosity Distance}
In $\Lambda$CDM, the luminosity distance of a Supernova at the heliocentric redshift $z_\text{helio}$ and the \ac{CMB} rest frame redshift $z_\text{CMB}$ is \cite{Akarsu_2019}
\begin{equation}
\begin{aligned}
    d_L &= (1+z_\text{helio})\int_0^{z_\text{CMB}} \frac{dz}{H(z)}\\
    &= (1+z_\text{helio}) \cdot \text{ Comoving Distance to }z_\text{CMB}.
\end{aligned}
\end{equation}
Extending this to the anisotropic universe is straightforward, by replacing the \ac{FLRW} comoving distance by $D_C$ defined in \cref{comoving_distance}.

\section{Methodology and Data\label{Method_data}}
\subsection{Methodology \label{Method}}
We numerically solve the thermodynamics evolution and the expansion of the universe using the LSODA method \cite{Hindmarsh_1983, Petzold_1983}.\footnote{Our code is available in Ref. \cite{Ng_2025}.} Notably, we use the Saha equation to solve for the electron fraction $X_e$ at the early times. Once the temperature cools to $\SI{6500}{\kelvin}$, we solve for $X_e(t)$, and thus, $t_*$ and $t_d$ by using the modified Peebles' formalism outlined in \cref{Thermo}. To account for the approximation in Peebles' formalism and possible numerics, a small correction factor is introduced to $t_*$ and $t_d$ based on a comparison with the standard cosmological code \texttt{CLASS} \cite{Blas_2011} alongside the recombination code \texttt{HyRec} \cite{HyRec-2, HyRec}. We also check for the accuracy of our code by comparing it with \texttt{CLASS} \cite{Blas_2011} under the standard $\Lambda$CDM setting. The discrepancies in both the photon last scattering redshift and $100\theta_*$ are within the \textit{Planck} error bars, indicating good agreement.

To search for suitable anisotropic cosmological parameters, we make use of the Monte Carlo code \texttt{Cobaya} \cite{Torrado_2019, Torrado_2021} with the built-in \ac{MCMC} sampler \cite{Lewis_2002, Lewis_2013}. A convergence threshold of Gelman-Rubin statistics quantity $R-1<0.01$ is chosen. We consider the parameter space $\{H_{x0}, \omega_{b0}, \omega_{c0},\alpha\}$, where
\begin{gather}
    \alpha \equiv \frac{H_{z0}}{H_{x0}}-1,\\
\intertext{and the physical densities}
    \omega_{b0} \equiv \frac{8\pi\rho_{b0}}{3\times(\SI{e-2}{\km\per\s\per\mega\pc})^2}\xrightarrow{\text{\ac{FLRW}}}\Omega_{b}h^2,\\
    \omega_{c0} \equiv \frac{8\pi\rho_{c0}}{3\times(\SI{e-2}{\km\per\s\per\mega\pc})^2}\xrightarrow{\text{\ac{FLRW}}}\Omega_{c}h^2,
\end{gather}
carrying similar meaning as $\Omega_{b} h^2$ and $\Omega_{c} h^2$ under the \ac{FLRW} metric, respectively. The priors are summarized in \cref{tab:prior}. In particular, only a relatively small $|\alpha|<\num{e-9}$ is considered since we require the present-day \ac{CMB} temperature anisotropies $\delta T/T\equiv|T_{x0}-T_{z0}|/\SI{2.7255}{\kelvin}<\num{2e-5}$ with
\begin{align}
    &T_{x0} = T(t_*)\cdot\frac{a(t_*)}{a_0}, &&T_{z0} = T(t_*)\cdot\frac{c(t_*)}{c_0}.
\end{align}
We emphasize that this prior choice is motivated by \ac{CMB} observations, specifically the \ac{CMB} temperature's fractional directional variation at the level of $<\num{e-5}$. Hence, $\alpha$ cannot be too large, as this would produce \ac{CMB} non-uniformity exceeding the observational limit. Moreover, since $\alpha$ can be positive or negative, corresponding to whether the $z$-axis is expanding faster than the $x$ and $y$-axes or not, we consider 2 different choices of the prior for $\alpha$, referred to as Prior \textit{A} and \textit{B}, respectively. We also explore a looser constraint, $\delta T/T <\num{e-3}$ to test whether larger anisotropies are favored. Therefore, we consider 2 additional sets of prior, Prior \textit{C} and \textit{D} with a wider range for $\alpha$. Finally, the results are analyzed using the Python package \texttt{GetDist} \cite{Lewis_2019} and a modified version of \texttt{MCEvidence} \cite{Heavens_2017}, specifically adapted for \texttt{Cobaya} chains.

\begin{table*}
\centering
\begin{tabularx}{\textwidth}
    {>{\raggedright\arraybackslash\hsize=.7\hsize\linewidth=\hsize}X
    >{\raggedright\arraybackslash\hsize=1.05\hsize\linewidth=\hsize}X
    >{\raggedright\arraybackslash\hsize=1.05\hsize\linewidth=\hsize}X
    >{\raggedright\arraybackslash\hsize=1.05\hsize\linewidth=\hsize}X
    >{\raggedright\arraybackslash\hsize=1.05\hsize\linewidth=\hsize}X
    >{\raggedright\arraybackslash\hsize=1.05\hsize\linewidth=\hsize}X}
    \toprule\toprule
    & & \multicolumn{2}{c}{$\delta T/T < \num{2e-5}$} & \multicolumn{2}{c}{$\delta T/T < \num{e-3}$} \\
    \cmidrule(lr){3-4}
    \cmidrule(lr){5-6}
    Parameter & $\Lambda$CDM & Prior \textit{A} & Prior \textit{B} & Prior \textit{C} & Prior \textit{D} \\\toprule
    $H_{x0}$ & $\mathcal{U}$(\num{60}, \num{80}) & $\mathcal{U}$(\num{60}, \num{80}) & $\mathcal{U}$(\num{60}, \num{80}) & $\mathcal{U}$(\num{60}, \num{80}) & $\mathcal{U}$(\num{60}, \num{80})\\
    $\omega_{b0}$ & $\mathcal{U}$(\num{0.005}, \num{0.1})& $\mathcal{U}$(\num{0.005}, \num{0.1}) & $\mathcal{U}$(\num{0.005}, \num{0.1}) & $\mathcal{U}$(\num{0.005}, \num{0.1}) & $\mathcal{U}$(\num{0.005}, \num{0.1}) \\
    $\omega_{c0}$ & $\mathcal{U}$(\num{0.001}, \num{0.99}) & $\mathcal{U}$(\num{0.001}, \num{0.99}) & $\mathcal{U}$(\num{0.001}, \num{0.99}) & $\mathcal{U}$(\num{0.001}, \num{0.99}) & $\mathcal{U}$(\num{0.001}, \num{0.99}) \\
    $\alpha$ & 0 & $\mathcal{U}$(\num{-1E-9}, \num{0}) & $\mathcal{U}$(\num{0}, \num{1E-9}) & $\mathcal{U}$(\num{-3E-8}, \num{0}) & $\mathcal{U}$(\num{0}, \num{3E-8})\\
    \bottomrule\bottomrule
\end{tabularx}
\caption{The priors used in this work. The four sets of anisotropic priors considered are referred to as \textit{A}, \textit{B}, \textit{C}, and \textit{D}. Here, $\mathcal{U}(x_1, x_2)$ refers to a uniform distribution over the interval $x_1$ to $x_2$.}
\label{tab:prior}
\end{table*}

\subsection{Data \label{Data}}
Due to the rapid isotropization \cite{Wald_1983,Akarsu_2019,Hertzberg_Loeb_2024}, only the measurements involving early time quantities have decent constraining power on the anisotropy. Hence, we are motivated to consider observations of the \ac{CMB} angular acoustic scale and the \ac{BAO} (inverse) angular acoustic scales. To supplement and to better constrain the standard cosmological parameters, $H(z)$ measurements from cosmic chronometer, Lyman-alpha, and \ac{BAO}, as well as magnitude measurements from \acp{SNIa} are employed.

Since \ac{CMB} and \ac{BAO} measurements already cover a large portion of the sky, full-sky averaged quantities approximate the actual observations well. The definition of full-sky observables is
\begin{gather}
    \langle \mathrm{Observable} \rangle = \frac{1}{2}\int_0^\pi (\mathrm{Observable}) \sin\Theta d\Theta.
\end{gather}
In contrast, Hubble parameter measurements and \ac{SNIa} measurements are often based on well-localized observations. Thus, it seems that we should consider their position information which would introduce three additional Euler angles to describe the orientation of the anisotropic axes relative to our astronomical coordinate system. However, owing to the rapid isotropization \cite{Wald_1983,Akarsu_2019, Hertzberg_Loeb_2024}, at late times, when Supernova and Hubble parameter measurements are based, $a_\mathrm{eff}\approx a\approx c$ within \num{e-8} levels. Therefore, there is virtually no anisotropy in these local observables, meaning that treating these observables as full-sky averaged quantities is a reasonable approximation, and it has the advantage of reducing the number of parameters by three. Explicitly, this work considers the averaged quantities $\langle H\rangle$ and $\langle d_L\rangle$ for Hubble parameter and \ac{SNIa} measurements, respectively.

In addition, we must develop a conversion between redshift and cosmological time since measurements are often given in redshift while our model is a function of cosmological time. For a given redshift $z_\mathrm{obs}$, the corresponding time $t_\mathrm{obs}$ should satisfy
\begin{gather}
    1+z_\mathrm{obs} = \frac{1}{\langle a_\mathrm{eff}\rangle(t_\mathrm{obs})},\\
\intertext{where}
    \langle a_\mathrm{eff}\rangle(t_\mathrm{obs})=\frac{1}{2}\int_{0}^{\pi}a_\mathrm{eff}(\Theta, t_\mathrm{obs})\sin\Theta d\Theta,
\end{gather}
is the full-sky averaged scale factor.

\subsubsection{\acs{CMB} Angular Acoustic Scale}
\textit{Planck} 2018 gives a highly precise measurement of the angular acoustic scale with $100\theta_{*,\text{ P18}}={1.04109}\pm{0.00030}$ (\textit{Planck} 2018  base\_plikHM\_TTTEEE\_lowl\_lowE) \cite{Planck_2018}. Although this value is obtained under the  $\Lambda$CDM model, the geometric nature of $\theta_*$ implies that it is rather independent of cosmology (see Sec. 3.1 and Table 5 in Ref. \cite{Planck_2018}). Thus, even for an anisotropic universe, this value serves as a good reference, and we can construct a Gaussian log-likelihood as
\begin{gather}
    \ln\mathcal{L}= -\frac{(100\langle\theta_{*,\text{th}}\rangle - 100\theta_{*,\text{P18}})^2}{2\sigma_\text{P18}^2},
\end{gather}
where $\sigma_\text{P18}=\num{0.00030}$ and the subscript `th' implies the theoretical value from the anisotropic model. 

\subsubsection{\acs{BAO} (Inverse) Angular Acoustic Scales}
This work considers the recent \textit{DESI} 2024 \ac{BAO} measurements \cite{DESI_2024lzq,DESI_2024mwx,DESI_2024uvr} with the \textit{SDSS} DR16 \cite{Alam_2020}, DR7 \cite{Ross_2014} and \textit{6dF} \cite{Beutler_2012} \ac{BAO} measurements. We adopt the \ac{BAO} likelihood code in \texttt{Cobaya} \cite{Torrado_2019, Torrado_2021} with slight modifications. We adjust the code to consider the full-sky averaged observables and to marginalize \ac{BAO} observables not listed in \cref{Angular_Scales}, such as $ f\sigma_8$.

\subsubsection{H(z) Measurements}
We follow Ref. \cite{Akarsu_2019} and evaluate our model under the same $H(z)$ dataset, namely a total of 36 Hubble parameter measurements from Cosmic Chronometer (CC) \cite{Moresco_2012,Zhang_2014,Simon_2005,Moresco_2016,Ratsimbazafy_2017,Moresco_2015}, Lyman-alpha (alone or with quasi-stellar objects) (Ly-$\alpha$) \cite{Delubac_2015,Font-Ribera_2014} and \ac{BAO} signals in galaxy distribution \cite{Alam_2017} (see Table II in Ref. \cite{Akarsu_2019} for details). We also adopt the $\chi^2$ definition in Ref. \cite{Akarsu_2019}, viz. (1) for the uncorrelated CC and Ly-$\alpha$ measurements
\begin{gather}
    \ln\mathcal{L}=\frac{\chi_{\text{CC/Ly-}\alpha}^2}{2} = \sum_{\text{CC/Ly-}\alpha} \frac{\Bigl(\langle H_\text{th}\rangle(z_i)-H_\text{obs}(z_i)\Bigr)^2}{2\sigma_\text{obs}^2(z_i)},
\end{gather}
and (2) for correlated galaxy distribution measurements
\begin{gather}
    \ln\mathcal{L}=\frac{\chi_{\text{galaxy}}^2}{2} = \frac{1}{2}D^T C^{-1} D.
\end{gather}
Here, $H_\text{obs}$ ($\langle H_\text{th}\rangle$) is the full-sky averaged Hubble parameter measured (predicted), $\sigma^2_\text{obs}$ is the square of the measurement error, $D$ is the vector of the differences between the predicted and measured values, and $C$ is the covariance matrix of the measurements.

\subsubsection{\acs{SNIa} Measurements}
The anisotropic model is analyzed with the Pantheon dataset \cite{Pantheon_2017} alongside the built-in likelihood code in \texttt{Cobaya} \cite{Torrado_2019, Torrado_2021} modified to take into account the full-sky averaged luminosity distance.

\section{Results and Discussion\label{Results}}
\begin{figure}
    \centering
    \includegraphics[width=\columnwidth, height=0.4\textheight, keepaspectratio]{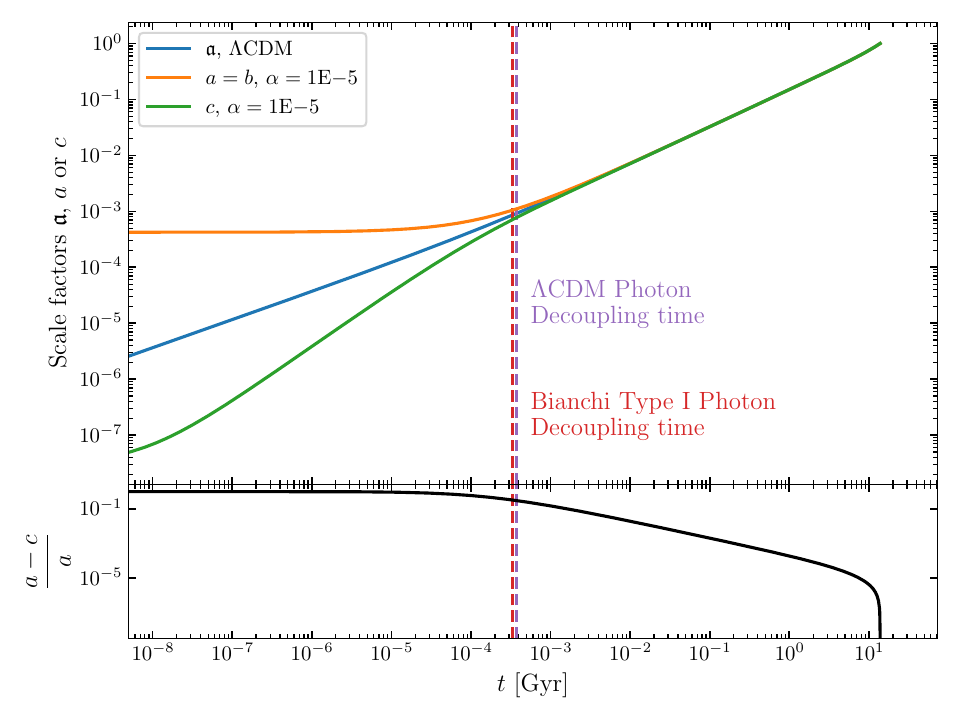}
    \includegraphics[width=\columnwidth, height=0.4\textheight, keepaspectratio]{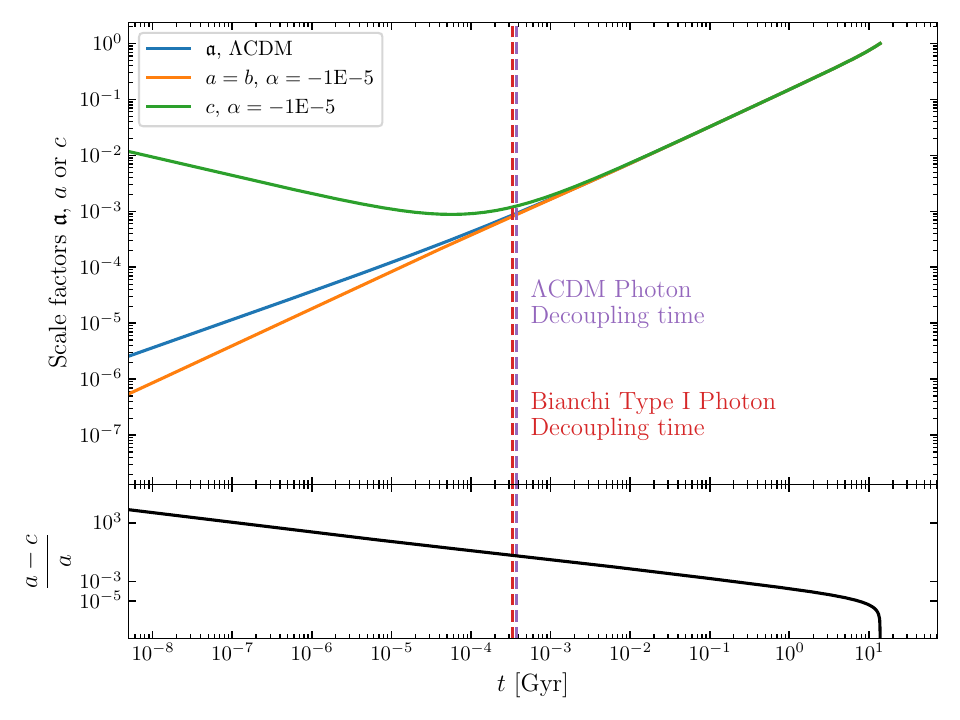}
    \caption{Expansion histories of $\Lambda$CDM and anisotropic universes with $\alpha=\pm10^{-5}$, $+$ ($-$) for the upper (bottom) figure. The bottom panel of each figure shows the fractional difference of the scale factors in the anisotropic universes. The photon decoupling time in each model is shown in the vertical dashed line. Here, $H_{x0}$ (or $\mathfrak{H}_0$ for $\Lambda$CDM) and the energy densities of all universes are from \textit{Planck} TTTEEE+low l+low E. \label{expansion_hist}}
\end{figure}

We would like to highlight the unique expansion history of a Bianchi Type I universe. Depending on which axis expands faster, the universe may undergo contraction along a direction initially and expansion in a later epoch, or the universe may start with a finite scale factor instead of zero. This is in agreement with Ref. \cite{Hertzberg_Loeb_2024} and is illustrated in \cref{expansion_hist} with exaggerated $\alpha$ variables. The asymmetry between positive and negative $\alpha$ is due to the \ac{LRS} nature. Positive $\alpha$ means two axes are expanding at a faster rate while negative $\alpha$ implies one axis expands faster than the rest. Another key property of the Bianchi Type I model is the rapid isotropization of the universe \cite{Wald_1983,Akarsu_2019,Hertzberg_Loeb_2024}. From \cref{expansion_hist}, we can see that even with exaggerated $\alpha$ variables, the anisotropy is still quickly washed out around the time of photon decoupling which happens at approximately the same time as $\Lambda$CDM. Hence, we would expect \ac{CMB} and \ac{BAO}, which probe physics at early times, to give the most stringent anisotropy constraint while local measurements are rather weak in detecting anisotropy.

\cref{Result_table} summarizes the \ac{MCMC} result as well as the log-Bayesian evidence and \ac{AIC}. Log-Bayesian evidence and \ac{AIC} are the model selection criteria, in particular 
\begin{gather}
    \text{AIC} = 2k - 2\ln\mathcal{L}_\text{max},
\end{gather}
where $k$ is the number of parameters and $\mathcal{L}_\text{max}$ is the maximum likelihood of the model.

\begin{figure}
    \centering
    \includegraphics[width=\columnwidth]{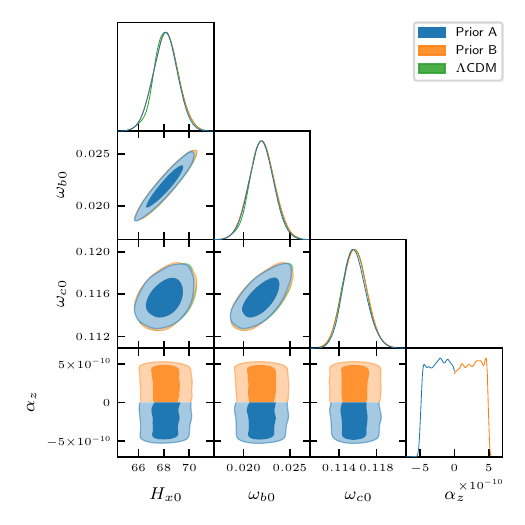}
    \caption{One and two-dimensional marginalized distributions (68\% and 95\% contours) of the cosmological parameters for the $\Lambda$CDM model and the anisotropic model with Priors \textit{A} and \textit{B}. \label{Result_fig_tight}}
\end{figure}

\begin{figure}
    \centering
    \includegraphics[width=\columnwidth]{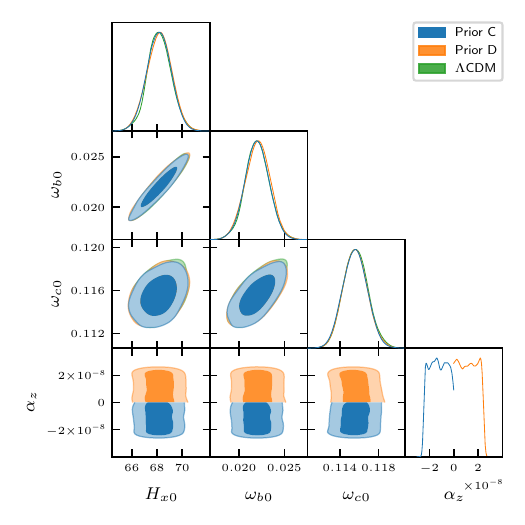}
    \caption{Same as \cref{Result_fig_tight}, but for Priors \textit{C} and \textit{D}. \label{Result_fig_mid}}
\end{figure}

From \cref{Result_table}, one can see that the mean and variance of the standard cosmological parameters ($H_{x0}$, $\omega_{b0}$, $\omega_{c0}$, $\omega_{m0}$) are consistent with the $\Lambda$CDM values for anisotropic priors. Moreover, the anisotropy parameter $\alpha$ is tightly constrained, with a 95\% limit of $>\num{-4.74e-10}$ for Prior \textit{A}. For Prior \textit{B}, while the posterior distribution shows a non-zero mean (\num{2.5e-10}) for $\alpha$, the $1\sigma$ uncertainty interval still encompasses $\alpha = 0$. When relaxing the maximum temperature anisotropy constraint to $\delta T/T < \num{e-3}$, these bounds become approximately 20 to 100 times weaker. A pictorial representation of the \ac{MCMC} result is shown in \cref{Result_fig_tight,Result_fig_mid} which plot the one and two-dimensional marginalized distributions (68\% and 95\% contours) of the parameters under different priors. We observe that $\alpha$ exhibits only weak correlations with standard cosmological parameters, while showing sharply bounded posterior distributions. The anisotropy parameter is directly related to $\delta T/T$, and excessively large values of $|\alpha|$ would violate the observational limits on $\delta T/T$. In contrast, standard cosmological parameters affect $\delta T/T$ only indirectly through their impact on the expansion history. This explains their weak correlation with $\alpha$. Importantly, the sharply bounded $\alpha$ posteriors reflect observational constraints from $\delta T/T$ rather than prior dominance.

Besides, \cref{Result_table} provides the constraints on $\tilde{\Omega}_{\sigma0}$ and $\delta_{xz}$ defined in Ref. \cite{Akarsu_2019} and \cite{Hertzberg_Loeb_2024}
\begin{gather}
    \tilde{\Omega}_{\sigma0}=\frac{(H_{x0}-H_{z0})^2}{(2H_{x0}+H_{z0})^2},\\
    \delta_{xz}=\frac{2|H_x(t_*)-H_z(t_*)|}{H_x(t_*)+H_z(t_*)},
\end{gather}
for comparison. We find that under the tight temperature anisotropy limit of \num{2e-5}, our result ($\tilde{\Omega}_{\sigma0}\lesssim\num{2.5e-20}$, $\delta_{xz}\lesssim\num{e-5}$) generally agrees with those in Ref. \cite{Akarsu_2019,Hertzberg_Loeb_2024} ($\tilde{\Omega}_{\sigma0}\lesssim\num{4e-20}$, $\delta_{xz}\lesssim\num{e-5}$, respectively).

\begin{table*}
\begin{center}

\begin{tabularx}{\textwidth}
    {>{\raggedright\arraybackslash}X
    >{\centering\arraybackslash}X
    >{\centering\arraybackslash}X
    >{\centering\arraybackslash}X
    >{\centering\arraybackslash}X
    >{\centering\arraybackslash}X}
\toprule\toprule & & \multicolumn{2}{c}{$\delta T/T < \num{5e-5}$} & \multicolumn{2}{c}{$\delta T/T < \num{e-3}$} \\
Parameter & $\Lambda$CDM & Prior \textit{A} & Prior \textit{B} & Prior \textit{C} & Prior \textit{D} \\
\toprule
\boldmath$H_{x0}         $ & $68.19\pm 0.95             $ & $68.11\pm 0.95             $ & $68.14\pm 0.98             $ & $68.15\pm 0.96             $ & $68.19\pm 0.99             $ \\
\boldmath$\omega_{b0}    $ & $0.0220\pm 0.0013          $ & $0.0219\pm 0.0013          $ & $0.0220\pm 0.0014          $ & $0.0220\pm 0.0013          $ & $0.0221\pm 0.0014          $ \\
\boldmath$\omega_{c0}    $ & $0.1156\pm 0.0013          $ & $0.1157^{+0.0011}_{-0.0013}$ & $0.1157\pm 0.0013          $ & $0.1156\pm 0.0013          $ & $0.1156\pm 0.0012          $ \\
{\boldmath$\alpha_z       $} (95\% C.I.) & --- & $> -4.73\cdot 10^{-10}     $ & $\left(\,2.5^{+2.2}_{-2.5}\,\right)\cdot 10^{-10}$ & $\left(\,-13^{+12}_{-11}\,\right)\cdot 10^{-9}$ & $< 2.37\cdot 10^{-8}       $ \\
$\omega_{m0}               $ & $0.1377\pm 0.0024          $ & $0.1376^{+0.0022}_{-0.0024}$ & $0.1377\pm 0.0024          $ & $0.1376\pm 0.0023          $ & $0.1377\pm 0.0024          $ \\
$\delta_{xz}               $ (95\% C.I.) & --- & $< 2.69\cdot 10^{-5}       $ & $\left(\,1.5^{+1.3}_{-1.4}\,\right)\cdot 10^{-5}$ & $0.00072^{+0.00065}_{-0.00069}$ & $< 0.00135                 $ \\
$\tilde{\Omega}_{\sigma0}  $ (95\% C.I.) & --- & $< 2.48\cdot 10^{-20}      $ & $< 2.54\cdot 10^{-20}      $ & $< 6.29\cdot 10^{-17}      $ & $< 6.26\cdot 10^{-17}      $ \\
\bottomrule \noalign{\vspace{2pt}}
$\ln\mathcal{E}$& -555.62 & -556.35 & -556.32 & -555.83 & -555.86 \\
AIC & $1090.48$ & $1092.48$ & $1092.46$ & $1092.46$ & $1092.44$ \\
\bottomrule\bottomrule
\end{tabularx}
\end{center}
\caption{Means and 68\% limits of the cosmological parameters corresponding to the $\Lambda$CDM model and the anisotropic priors under different maximum temperature anisotropy $\delta T/T$. For $\alpha$, $\tilde{\Omega}_{\sigma0}$ and $\delta_{xz}$, the 95\% upper limits are shown. The log-Bayesian evidence and \ac{AIC} are also listed. Note that for Priors \textit{A} and \textit{C}, $\alpha$ is negative. \label{Result_table}}
\end{table*}

To determine whether the anisotropic model is preferred over $\Lambda$CDM, one would need to consider the Bayes factor or \ac{AIC}. The Bayes factor $\mathcal{B}$ is defined as the ratio of the evidence of the model under consideration to that of the fiducial model \cite{Kass_1995}. A ratio larger than 1 implies the model under consideration is favored over the fiducial \cite{Kass_1995}. In contrary, a positive $\Delta\text{AIC}\equiv\text{AIC}_\text{model}-\text{AIC}_\text{fiducial}$ implies the model under consideration is disfavored \cite{Burnham_1998}. We find that regardless of the prior choice, the \ac{LRS} Bianchi Type I spacetime extension of the $\Lambda$CDM model is consistently not statistically preferred compared to $\Lambda$CDM, with a log-Bayes factor $<\num{1}$ and a $\Delta$\ac{AIC} $<\num{2}$. The lack of preference over $\Lambda$CDM mainly stems from the fact that the $\Lambda$CDM can produce a full-sky averaged angular acoustic scale $\langle\theta_*\rangle$ as good as the \ac{LRS} Bianchi Type I, within \textit{Planck} error bars. We expect that the anisotropic model might be preferred over $\Lambda$CDM to a large extent if the $\theta_*$ at each direction is studied instead of taking the full sky average. Observations are already able to detect anisotropy in $100\theta_\text{MC}$, which is approximately $100\theta_*$, by dividing the sky into \num{48} half-skies \cite{Yeung_2022}. It would be interesting to follow up on this direction in the future.

\begin{table}
    \centering
    \begin{tabularx}{\columnwidth}
        {>{\centering\arraybackslash\hsize=.6\hsize\linewidth=\hsize}X
        >{\raggedright\arraybackslash\hsize=1.4\hsize\linewidth=\hsize}X}
    \toprule\toprule
    $\ln\mathcal{B}$ & Evidence against fiducial\\
    0 to 1 & Not worth a bare mention\\
    1 to 3 & Positive evidence\\
    3 to 5 & Strong evidence\\
    $>5$ & Very strong evidence\\
    \bottomrule\toprule
    $\Delta\text{AIC}$ & Evidence for fiducial\\
    0 to 2 & Substantial evidence\\
    4 to 7 & Considerably less evidence\\
    $>10$ & Essentially no evidence\\
    \bottomrule\bottomrule
    \end{tabularx}
    \caption{Jeffery's scale for Bayes factor \cite{Kass_1995} alongside the correspondence between $\Delta$\ac{AIC} and evidence for the fiducial \cite{Burnham_1998}. \label{Bayes_AIC}}
\end{table}

In the following, we discuss the significance of the modifications in the definitions of observables and the recombination history. First, it must be stressed that the proper way to define observables is through considering geodesics. One should not take the definitions in the \ac{FLRW} metric and promote the \ac{FLRW} scale factor $\mathfrak{a}$ to the geometric averaged scale factor $\tilde{a}$ without any justification. Although the definitions agree with each other by reducing to the standard ones in the \ac{FLRW} metric when anisotropy could be negligible, it is important to distinguish them especially when the anisotropy is large, suggested by some observations such as those in Ref. \cite{Yeung_2022}. For instance, we consider the means of parameters in Prior \textit{A} and \textit{C} and study how the \ac{CMB} angular acoustic scales differ. In Prior \textit{A}, the difference in $100\langle\theta_*\rangle$ between the definitions is only around \num{7e-3}\%, substantially smaller than the percentage error of \textit{Planck} (0.03\%)\cite{Planck_2018}. However, in Prior \textit{C} which is 50 times more anisotropic than the former, the difference in $100\langle\theta_*\rangle$ grows to a considerable level of 0.2\%, an order larger than the error of \textit{Planck}.

In contrast, the modified recombination history has little to no impact on the last scattering time $t_*$. Our parameter means from \ac{MCMC} suggest that the last scattering time $t_*$ differ by $<0.01\%$ in all prior sets, compared to our $\Lambda$CDM \ac{MCMC} result, $t_*\sim\SI{3.79E-4}{\giga\year}$. Moreover, the corresponding redshifts as defined in \cref{Akarsu_redshift} are consistent with that from our $\Lambda$CDM result, $\tilde{z}_*\sim1085$. Thus, under realistic anisotropy, one can safely ignore the change in the recombination history of the Bianchi Type I model. Since the thermodynamics are not significantly changed, we redo the \ac{MCMC} analysis for $\delta T/T<\num{2E-3}$ and the $\Lambda$CDM case using the codes \texttt{HyRec} \cite{HyRec,HyRec-2} and modified \texttt{CLASS} \cite{Blas_2011}, based on the equations prescribed in \cref{GR} and Ref. \cite{Akarsu_2019} for a more sophisticated atomic treatment. The results, presented in \cref{MCMC_HR}, are generally consistent with \cref{Result_table}, indicating our thermodynamics treatment which follows Peebles' three-level atom model, is decent. We also point out that since there is no strong correlation between $\alpha$ and $\omega_{b0}$ for the range of anisotropy we considered, varying baryon density only introduces larger error bars to the standard cosmological parameters.

This work is motivated by the anisotropy observed in global and local measurements, particularly directional dependencies in the Hubble constant \cite{Yeung_2022, Javanmardi_2015, Krishnan_2022, Luongo_2022, Hu_2023}. A meaningful comparison can be made between our constraints and the observed anisotropies. Generally, local observations on \acp{SNIa} \cite{Javanmardi_2015, Krishnan_2022, Luongo_2022, Hu_2023} find that the Hubble constant varies by $\sim\SI{1}{\km\per\s\per\mega\pc}$ with direction, equivalent to a fractional difference of $\sim1\%$. Our result, on the other hand, suggests that the fractional difference is no more than $\sim\num{e-6}$\% even when considering the looser limits, i.e., results from Priors \textit{C} and \textit{D}. Hence, the Bianchi Type I model is definitely inadequate in explaining the observed \acp{SNIa} anisotropy, which is not surprising. Recall that the Bianchi Type I model isotropizes at late times. To achieve a \SI{1}{\km\per\s\per\mega\pc} variation in the Hubble constant, the universe must be highly anisotropic at late times and even more so at early times, disrupting physics at the last scattering surface. Therefore, the observed anisotropy in \acp{SNIa} needs a model that anisotropize at late times to explain, such as through the presence of anisotropic matter sources \cite{Koivisto_2008}. Alternatively, the anisotropy may just be due to local effects, but it is challenging to explain its alignment with \ac{CMB} multipoles.

\begin{figure}
    \centering
    \includegraphics[width=\columnwidth]{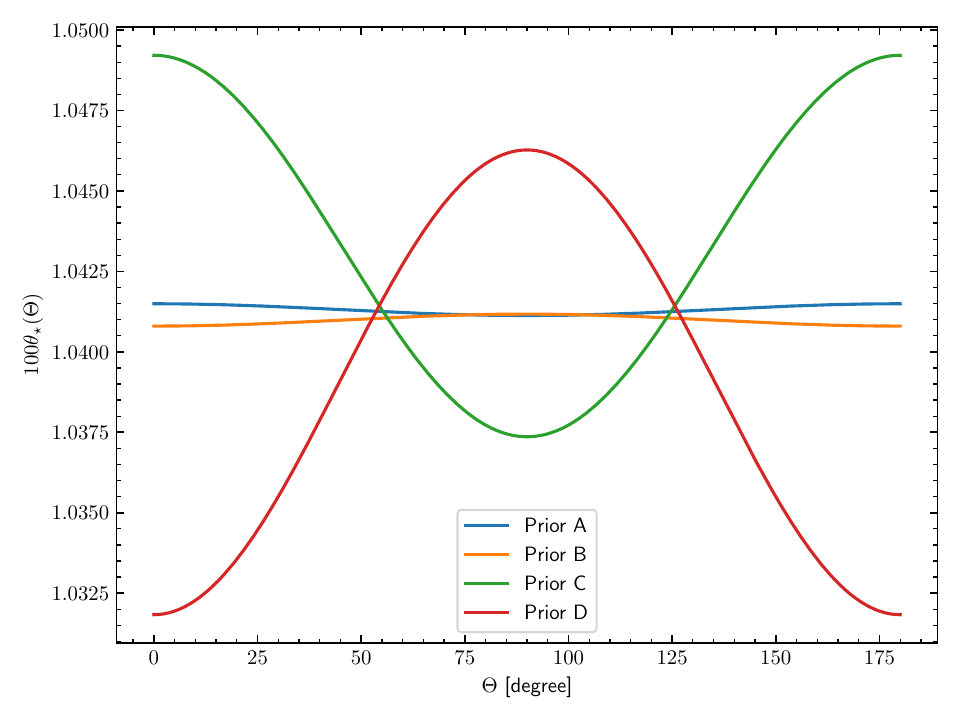}
    \caption{Directional variation of $100\theta_*$ from mean values of different priors. \label{theta_variation}}
\end{figure}

Comparing our results with the observed \ac{CMB} anisotropies in $\{\Omega_{b} h^2, \Omega_{c} h^2, n_s, 100\theta_\mathrm{MC}, \tau,\allowbreak \ln{(10^{10}A_s)},\allowbreak H_0\}$ is challenging, as these are derived from \ac{MCMC} analyzes of \textit{Planck} half-skies power spectra with the standard $\Lambda$CDM cosmology \cite{Yeung_2022}. If the universe is indeed anisotropic, the \ac{MCMC} result obtained using the $\Lambda$CDM model is inconsistent and, thus, not representative. Regardless, one can still compare our result with Ref. \cite{Yeung_2022}. Clearly, our model assumes that energy densities have no directional variation which disagrees with Ref. \cite{Yeung_2022}. Moreover, Ref. \cite{Yeung_2022} observed a variation in $H_0$ with amplitude $\sim\SI{5}{\km\per\s\per\mega\pc}$. Again, such a large variation is unreachable in our model. However, it may be more important to compare the anisotropy in $100\theta_\mathrm{MC}$ due to its geometric nature and robustness to different cosmologies \cite{Planck_2018}. The large directional variations in Hubble constant and energy densities may be the consequence of fitting the standard $\Lambda$CDM cosmology to the observed $100\theta_\mathrm{MC}$ anisotropy instead. From \cref{theta_variation}, we can see that Prior \textit{A} (\textit{B}) predicts a variation roughly between \numrange{1.0411}{1.0415} (\numrange{1.0408}{1.0412}) while it is between \numrange{1.0374} {1.0492} (\numrange{1.0318}{1.0463}) for Prior \textit{C} (\textit{D}). To compare, Ref. \cite{Yeung_2022} reports a fluctuation from \numrange{1.0391}{1.0423}, which is larger than that from Prior \textit{A} and \textit{B}, but smaller than that from Prior \textit{C} and \textit{D}. Hence, if we relax the temperature anisotropy constraint, our model can explain the amplitude of the observed anisotropy in $100\theta_\mathrm{MC}$. Yet, our \ac{LRS} model cannot explain the azimuthal dependency observed \cite{Yeung_2022}. It would be our next focus to study such anisotropy in $100\theta_\mathrm{MC}$ with the fully asymmetry Bianchi Type I model.

To further improve the constraints on the anisotropic model with currently available data, one needs to develop the theory and cosmological codes for \ac{BBN} and the \ac{CMB} power spectra under the Bianchi Type I model. Current \ac{BBN} and cosmological codes such as \texttt{PArthENoPE} and \texttt{CLASS} \cite{Parthenope_2018, Blas_2011} assume the \ac{FLRW} metric. Significant work is needed to modify the codes for the Bianchi Type I metric, especially for \texttt{CLASS} which adopts \ac{FLRW} scale factor $\mathfrak{a}$ as the temporal coordinate. For an order-of-magnitude constraint on $\tilde{\Omega}_{\sigma0}$ and $\delta_{xy}$, one can refer to Ref. \cite{Akarsu_2019,Hertzberg_Loeb_2024,Akarsu_2023} since our result is consistent with theirs under a maximum temperature anisotropy $\delta T/T$ of \num{2e-5}. However, their analysis should be taken with caution since the authors assume that the \ac{BBN} time or redshift is not affected by the anisotropy \cite{Akarsu_2019,Hertzberg_Loeb_2024,Akarsu_2023}. Without proper study, the possibility of a drastically modified \ac{BBN} history with similar element fractions cannot be ruled out, especially since \ac{BBN} does not have a strong implication on isotropy like \ac{CMB}. Meanwhile, the constraints can be improved if one detects primordial relics such as primordial neutrinos. Their energy would have directional dependency due to the anisotropic dilution after freeze out \cite{Hertzberg_Loeb_2024}. Contrarily, if the \ac{CMB} dipole is found to have a non-zero kinematic contribution, it would be a strong piece of evidence for the anisotropic universe and possibly for the Bianchi Type I model. Rigorous studies on \ac{BBN}, which occurred at a more anisotropic time than \ac{CMB}'s, would be particularly valuable. The ongoing \textit{Euclid} mission and the future \textit{Roman} mission could provide crucial tests on the kinematic nature of \ac{CMB} by measuring the \ac{CIB} dipole \cite{Kashlinsky_2022,Euclid_2024,Roman_2019}.

\section{Conclusion\label{Conclusion}}
In this paper, we present a self-consistent constraint on the \ac{LRS} Bianchi Type I spacetime extension of the $\Lambda$CDM model. Instead of assuming a fixed recombination redshift or time, we calculate the recombination history with a modified Peebles' formalism based on the anisotropic model. Moreover, we improve on previous works \cite{Akarsu_2019,Sarmah_2022, Yadav_2023, Koussour_2023, Akarsu_2023} by constructing distances and observables, including \ac{CMB} and \ac{BAO} angular acoustic scales, based on geodesic, achieving truly directional dependent observables. To constrain the model, we construct and modify existing likelihood codes to compare the full-sky averaged observables with observations in companion with \ac{MCMC}. Our key conclusions are summarized below:
\begin{enumerate}[(i)]
    \item By considering full-sky averages, the \ac{LRS} Bianchi Type I model is not statistically preferred compared to the $\Lambda$CDM model for maximum allowed temperature anisotropy $\delta T/T<\num{2e-5}$ and $\num{e-3}$.
    \item The anisotropy parameter $\alpha$ is tightly constrained with an upper bound in $|\alpha|$ of $\sim\num{e-10}$ (\num{e-8}) at $95\%$ confidence level when the maximum temperature anisotropy $\delta T/T$ equals \num{2e-5} (\num{e-3}). The tighter constraint is generally consistent with Ref. \cite{Akarsu_2019,Hertzberg_Loeb_2024} obtained using similar temperature anisotropy.
    \item We stress the importance of geodesic-based observables in anisotropic cosmology by showing that the geodesic-based \ac{CMB} angular scale is drastically different from that defined by drawing simple parallels with $\Lambda$CDM \cite{Akarsu_2019,Sarmah_2022, Yadav_2023, Koussour_2023, Akarsu_2023} under a non-negligible anisotropy.
    \item The modifications in the recombination history are insignificant, even under a large anisotropy such that maximum temperature anisotropy $\delta T/T<\num{e-3}$. 
    \item The anisotropic model cannot account for the observed directional dependency in \acp{SNIa} observations \cite{Javanmardi_2015, Krishnan_2022, Luongo_2022, Hu_2023}. On the other hand, the magnitude of the $100\theta_\mathrm{MC}$ anisotropy observed in \ac{CMB} \cite{Yeung_2022} can be explained with our model with a maximum temperature anisotropy $\delta T/T$ slightly larger than \num{2e-5}.
\end{enumerate}

\begin{acknowledgments}
    The computational resources used for the simulations in this work were kindly provided by The Chinese University of Hong Kong Central Research Computing Cluster. This research is supported by grants from the Research Grants Council of the Hong Kong Special Administrative Region, China, under project Nos. AoE/P-404/18 and 14300223. The Python packages \texttt{Numpy} \cite{Numpy_2020}, \texttt{Scipy} \cite{Scipy_2020}, \texttt{Numba} \cite{Numba_2015} and \texttt{Matplotlib} \cite{Matplotlib_2007} are used in this work.
\end{acknowledgments}

\appendix
\section{MCMC Results with CLASS Thermodynamics \label{MCMC_HR}}
Since Peebles' formalism neglects most atomic transitions and considers only a three-level Hydrogen atom \cite{Peebles_1968,Baumann_2022,Weinberg_2008}, it is crucial to verify that our results are consistent with those obtained from a more accurate thermodynamics code that incorporates contributions from Helium and a multi-level atomic treatment. As demonstrated in \cref{Results}, the thermodynamics remain virtually unchanged compared to the $\Lambda$CDM framework. Consequently, we employ the recombination code \texttt{HyRec} \cite{HyRec,HyRec-2} alongside a modified version of the cosmological code \texttt{CLASS} \cite{Blas_2011} according to \cref{GR} and Ref. \cite{Akarsu_2019}. We then perform \ac{MCMC} analyses on both the $\Lambda$CDM model and the anisotropic model, with the condition $\delta T/T<\num{2e-5}$. The results, which are consistent with those derived using Peebles' formalism, are presented in \cref{Result_table_HR}. This confirms that our findings are robust against more sophisticated treatments of thermodynamics and recombination.

\begin{table*}
\begin{center}

\begin{tabularx}{\textwidth}
    {>{\raggedright\arraybackslash}X
    >{\centering\arraybackslash}X
    >{\centering\arraybackslash}X
    >{\centering\arraybackslash}X
    >{\centering\arraybackslash}X
    >{\centering\arraybackslash}X
    >{\centering\arraybackslash}X}
\toprule\toprule & \multicolumn{3}{c}{Peebles' formalism} & \multicolumn{3}{c}{\texttt{HyRec} + modified \texttt{CLASS}} \\
\cmidrule(lr){2-4}
\cmidrule(lr){5-7}
Parameter & $\Lambda$CDM & Prior \textit{A} & Prior \textit{B} & $\Lambda$CDM HR & Prior \textit{A} HR & Prior \textit{B} HR \\
\toprule
\boldmath$H_{x0}         $ & $68.19\pm 0.95             $ & $68.11\pm 0.95             $ & $68.14\pm 0.98             $ & $68.20\pm 0.97             $ & $68.19\pm 0.96             $ & $68.16\pm 0.98             $ \\
\boldmath$\omega_{b0}    $ & $0.0220\pm 0.0013          $ & $0.0219\pm 0.0013          $ & $0.0220\pm 0.0014          $ & $0.0220\pm 0.0014          $ & $0.0220\pm 0.0014          $ & $0.0220\pm 0.0014          $ \\
\boldmath$\omega_{c0}    $ & $0.1156\pm 0.0013          $ & $0.1157^{+0.0011}_{-0.0013}$ & $0.1157\pm 0.0013          $ & $0.1157\pm 0.0011          $ & $0.1156\pm 0.0011          $ & $0.1157\pm 0.0011          $ \\
{\boldmath$\alpha_z       $} (95\% C.I.) & --- & $> -4.73\cdot 10^{-10}     $ & $\left(\,2.5^{+2.2}_{-2.5}\,\right)\cdot 10^{-10}$ & --- & $> -4.75\cdot 10^{-10}     $ & $< 4.74\cdot 10^{-10}      $ \\
$\omega_{m0}               $ & $0.1377\pm 0.0024          $ & $0.1376^{+0.0022}_{-0.0024}$ & $0.1377\pm 0.0024          $ & $0.1377\pm 0.0022          $ & $0.1376\pm 0.0022          $ & $0.1377\pm 0.0022          $ \\
$\delta_{xz}               $ (95\% C.I.) & --- & $< 2.69\cdot 10^{-5}       $ & $\left(\,1.5^{+1.3}_{-1.4}\,\right)\cdot 10^{-5}$ & --- & $< 2.71\cdot 10^{-5}       $ & $< 2.70\cdot 10^{-5}       $ \\
$\tilde{\Omega}_{\sigma0}  $ (95\% C.I.) & --- & $< 2.48\cdot 10^{-20}      $ & $< 2.54\cdot 10^{-20}      $ & --- & $< 2.50\cdot 10^{-20}      $ & $< 2.49\cdot 10^{-20}      $ \\
\bottomrule \noalign{\vspace{2pt}}
$\ln\mathcal{E}$& -555.62 & -556.35 & -556.32 & -555.52 & -556.27 & -556.26\\
AIC & $1090.48$ & $1092.46$ & $1092.45$ & $1090.45$ & $1092.46$ & $1092.45$ \\
\bottomrule\bottomrule
\end{tabularx}
\end{center}
\caption{Similar to \cref{Result_table}, but comparing \ac{MCMC} results from Peebles' recombination formalism and \texttt{HyRec} with the modified \texttt{CLASS}. `HR' here refers to the cases with \texttt{HyRec} and modified \texttt{CLASS}.\label{Result_table_HR}}
\end{table*}

\bibliography{Reference}
\end{document}